\documentclass[aps,prd,onecolumn,showkeys,groupedaddress,showpacs,nofootinbib]{revtex4}
\usepackage{epsfig} \usepackage{amsmath} \usepackage{amsfonts}
\usepackage{amssymb} \usepackage{graphicx} \usepackage{colordvi}
\usepackage{psfrag}
 \usepackage{times}
 \usepackage{amsmath, amsthm, amssymb}
 \usepackage{makeidx}

\newtheorem{Proposition}{Proposition}

\newfont{\gotico}{eufm10 scaled\magstephalf}
\newfont{\qvd}{msam10 scaled\magstephalf}

\def\de#1/de#2{\frac{\partial {#1}}{\partial {#2}}}
\def\De#1/de#2{\dfrac{\partial {#1}}{\partial {#2}}}

\def\const{{\rm const.}}

\makeindex \makeatletter
\def\widebar{\accentset{{\cc@style\underline{\mskip10mu}}}}
\makeatother

\begin{document}
%\numberwithin{equation}{section}
\def\bib#1{[{\ref{#1}}]}
\title{\bf The Cauchy problem for metric-affine $f(R)$-gravity in presence of a Klein-Gordon scalar field}

\author{S. Capozziello$^{1}$ and S. Vignolo$^{2}$ }

\affiliation{$~^{1}$ Dipartimento di Scienze Fisiche,
Universit\`{a} ``Federico II'' di Napoli and INFN Sez. di Napoli,
Compl. Univ. Monte S. Angelo Ed. N, via Cinthia, I- 80126 Napoli
(Italy)}

\affiliation{$^{2}$DIPTEM Sez. Metodi e Modelli Matematici,
Universit\`a di Genova,  Piazzale Kennedy, Pad. D - 16129 Genova
(Italy)}

\date{\today}

\begin{abstract}
We study  the initial value formulation of metric-affine $f(R)$-gravity  in
presence of a Klein-Gordon scalar field acting as source of the field equations. 
Sufficient conditions for the well-posedness of the Cauchy problem are  formulated. 
This result completes the analysis of the same problem already considered for other sources.

\end{abstract}

\pacs{04.50.+h, 04.20.Ex, 04.20.Cv, 98.80.Jr}
\keywords{Alternative theories of gravity; metric-affine approach;
initial value formulation }

\maketitle
\section{Introduction}
Among the various attempts to extend General Relativity, $f(R)$-gravity can
be considered a useful paradigm capable of preserving the good and well-established results of the 
Einstein gravity without imposing {\it a priori} the form of the action.  This approach has recently acquired
great interest in cosmology and in quantum field theory in order
to cure  the shortcomings of  standard theory of gravity at ultra-violet
and infra-red scales. In particular, $f(R)$-gravity, not excluded by observations \cite{nature}
could result useful   to address
cosmological puzzles as  dark energy
and dark matter, up to now not probed  at fundamental scales. On the other hand,  observations and experiments
could, in principle, help to reconstruct the effective form of the
theory by solving inconsistencies and shortcomings present at  various
scales in General Relativity (for reviews, see  \cite{OdiRev,GRGrew,faraonirev,defelice}).   

However, because of further gravitational degrees of freedom, emerging in assuming actions not linear in the Ricci scalar $R$, 
the initial value problem becomes a urgent issue to be  correctly formulated and addressed. Furthermore, also if the initial value problem is
well-formulated,   stability against perturbations has to be studied  and the causal structure has to be preserved.  If both these requirements are satisfied, the  initial value problem of the
theory is  also well-posed.

The Cauchy problem for General Relativity is well-formulated and well-posed  \cite{yvonne,yvonne2,Synge,Wald}.   Such a
result should be achieved also for  $f(R)$-gravity, if one wants to consider it a viable  extension of the Einstein
theory and compare the two approaches.

The crucial point with respect to General Relativity  comes from the fact that  the further
degrees of freedom could lead to an ill-formulated initial value
problem. In fact, their  role has to be clearly understood  in
order to discriminate among ghost modes, standard massless modes
or further massive gravitational modes \cite{veltman}.  Such an analysis  can be correctly addressed
by conformally transforming  $f(R)$-gravity  from the Jordan frame to the
Einstein frame. The extra-degrees of freedom give rise to  auxiliary
scalar fields, minimally coupled to the standard gravitational
Hilbert-Einstein action.  In this way, the analysis  results simplified.

However, being $f(R)$-gravity
a gauge theory, the initial  value formulation  depends on
suitable constraints and  gauges that mean a consequent
choice of coordinates so that the Cauchy problem can result
well-formulated and, possibly, well-posed. 
The debate on the well-formulation and the
well-posedness of the Cauchy problem of $f(R)$ theories, in metric and 
Palatini approaches,  has  recently  given several interesting and sometime contrasting results
\cite{Faraoni,CV1,CV2,CV3}.

It is possible to
show that the Cauchy problem of metric-affine $f(R)$-gravity is
well-formulated and well-posed in vacuo, while it can be, at
least, well-formulated for various form of matter fields \cite{CV3}.  The
reason of the apparent contradiction with respect to the results
in \cite{Faraoni} lies on the above mentioned gauge choice.
Following \cite{Synge}, Gaussian normal coordinates can be
adopted. This choice, introducing further constraints on the
Cauchy data surface, results more suitable to set the initial
value problem in such a way that the well-formulation can be
easily achieved. As a general remark, we can say that the
well-position cannot be achieved  for any metric-affine
$f(R)$-gravity theory but it has to be formulated specifying, case
by case, the source term in the field equations. However, it is
straightforward to demonstrate that, in vacuum case, as well as
for electromagnetic and generic Yang-Mills fields acting as
sources, the Cauchy problem results always well-formulated and
well-posed since it is possible to show that
$f(R)$-gravity reduces to $R+\Lambda$, that is the General
Relativity plus  a cosmological constant \cite{CV3}.

In \cite{CV1}, we have addressed the Cauchy problem for
metric-affine $f(R)$-gravity, in the Palatini approach and with
torsion, assuming  perfect-fluid matter as source. Performing the
conformal transformation from the Jordan to the Einstein frame and
following the approach by Bruhat, adopted for General Relativity
\cite{yvonne,yvonne2,yvonne4}, we have formulated sufficient conditions 
to ensure the well-posedness of the Cauchy problem. Moreover, we have shown 
that the set of functions $f(R)$ satisfying the stated conditions is actually not empty.

Here, we want to show that analogous results hold also in the case in which the source is a Klein-Gordon scalar field.
This is a delicate case due to the fact that second-order partial derivatives in the Klein-Gordon equation could give rise to 
inconsistencies when coupled with gravitational degrees of freedom (for a detailed discussion see \cite{Wald}).

The layout of the paper is the following. In Sec. II, we give a summary of $f(R)$-gravity in 
metric-affine formulation   {\it \`a la\/} Palatini and with torsion. Sec.III
is devoted to the discussion of the Cauchy
problem in presence of a Klein-Gordon scalar field acting as source of the field equations. We use the arguments in \cite{yvonne,yvonne2,Leray,yvonne4}  to show the consistency with the analogous  well-formulation and well-position of General Relativity.  In particular, it is possible to show that the well-position conditions are capable of selecting self-consistent $f(R)$-models.  The relevant example $f(R)=R+\alpha R^2$ is discussed in Sec.IV. Conclusions are drawn in Sec.V.

\section{Preliminaries on metric-affine $f(R)\/$-gravity}

The pairs $(g,\Gamma)\/$ constitute the gravitational fields  in the metric--affine formulation of $f(R)$-theories of gravity:
 $g$ is a pseudo-Riemannian metric  and $\Gamma$  a linear
connection  on the space-time manifold ${\cal M}$. 
The connection
$\Gamma\/$ is torsionless in the Palatini approach but it is not requested to be
metric--compatible. On the other hand,  in the approach with torsion, the
dynamical connection $\Gamma$ is forced to be metric in the approach with torsion.
The field equations are derived from the action
\begin{equation}\label{2.1}
{\cal A}\/(g,\Gamma)=\int{\left(\sqrt{|g|}f\/(R) + {\cal L}_m\right)\,ds}
\end{equation}
where $f(R)$ is a real function, $R\/(g,\Gamma) = g^{ij}R_{ij}\/$
(with $R_{ij}:= R^h_{\;\;ihj}\/$)  is the  curvature scalar
associated with the connection $\Gamma\/$ and ${\cal L}_m\,ds\/$
is the matter Lagrangian.

Assuming that the matter Lagrangian does not depend on the dynamical connection, the field equations are
\begin{subequations}\label{2.2}
\begin{equation}\label{2.2a}
f'\/(R)R_{ij} - \frac{1}{2}f\/(R)g_{ij}=\Sigma_{ij}\,,
\end{equation}
\begin{equation}\label{2.2b}
T_{ij}^{\;\;\;h} = - \frac{1}{2f'\/(R)}\de{f'\/(R)}/de{x^p}\/\left(\delta^p_i\delta^h_j - \delta^p_j\delta^h_i\right)
\end{equation}
\end{subequations}
for $f(R)$-gravity with torsion \cite{CCSV1}, and
\begin{subequations}\label{2.3}
\begin{equation}\label{2.3a}
f'\/(R)R_{ij} - \frac{1}{2}f\/(R)g_{ij}=\Sigma_{ij}\,,
\end{equation}
\begin{equation}\label{2.3b}
\nabla_k\/(f'(R)g_{ij})=0\,,
\end{equation}
\end{subequations}
for $f(R)$-gravity in the  Palatini approach \cite{francaviglia1,francaviglia2,Sotiriou,Sotiriou-Liberati1,Olmo}.  The quantity ${\displaystyle
\Sigma_{ij}:= -
\frac{1}{\sqrt{|g|}}\frac{\delta{\cal L}_m}{\delta g^{ij}}\/}$ is
the stress-energy tensor. From Eqs.
\eqref{2.2a} and \eqref{2.3a}, we obtain the relation
\begin{equation}\label{2.4}
f'\/(R)R -2f\/(R) = \Sigma
\end{equation}
where the curvature scalar $R\/$ is linked to  the trace of the stress-energy tensor
$\Sigma:=g^{ij}\Sigma_{ij}\/$.
From now on, we shall suppose that the relation \eqref{2.4}  is
invertible as well as that $\Sigma\not=\const\/$ (this implies  $f(R) \neq \alpha R^2\/$ which is 
compatible with $\Sigma=0\/$). Under these restrictions, the
curvature scalar $R\/$ can be expressed as a suitable function of
$\Sigma\/$, namely
\begin{equation}\label{2.5}
R=F(\Sigma)\,.
\end{equation}
If $\Sigma=\const$, General Relativity plus the cosmological constant is immediately recovered \cite{CCSV1}.
Defining the scalar field
\begin{equation}\label{2.6}
\varphi:=f'(F(\Sigma))\,,
\end{equation}
we can put the Einstein--like field equations of both {\it \`a
la\/} Palatini  and with torsion theories in the same form
\cite{CCSV1,Olmo}, that is
\begin{equation}\label{2.7}
\begin{split}
\tilde{R}_{ij} -\frac{1}{2}\tilde{R}g_{ij}= \frac{1}{\varphi}\Sigma_{ij}
+ \frac{1}{\varphi^2}\left( - \frac{3}{2}\de\varphi/de{x^i}\de\varphi/de{x^j}
+ \varphi\tilde{\nabla}_{j}\de\varphi/de{x^i} + \frac{3}{4}\de\varphi/de{x^h}\de\varphi/de{x^k}g^{hk}g_{ij} \right. \\
\left. - \varphi\tilde{\nabla}^h\de\varphi/de{x^h}g_{ij} -
V\/(\varphi)g_{ij} \right)\,,
\end{split}
\end{equation}
where  the effective potential
\bigskip\noindent
\begin{equation}\label{2.8}
V\/(\varphi):= \frac{1}{4}\left[ \varphi
F^{-1}\/((f')^{-1}\/(\varphi)) +
\varphi^2\/(f')^{-1}\/(\varphi)\right]\,,
\end{equation}
for the scalar field $\varphi\/$ has been introduced. In Eq. \eqref{2.7}, $\tilde{R}_{ij}\/$, $\tilde{R}\/$ and $\tilde\nabla\/$  
denote, respectively, the Ricci tensor, the scalar curvature and the covariant derivative associated with the Levi-Civita connection of the metric $g_{ij}$.
Therefore, if the dynamical connection $\Gamma\/$ is not coupled
with matter, both the theories (with torsion and {\it \`a
la\/} Palatini)
generate identical Einstein-like field equations. Moreover, it can be shown \cite{CCSV1} that the Einstein--like Eqs. \eqref{2.7} (together with Eqs. \eqref{2.6}) are deducible from a scalar-tensor theory with Brans-Dicke parameter $\omega_0=-3/2\/$. To see this point, we recall that the action functional of a (purely metric) scalar--tensor theory is given by
\begin{equation}\label{00001}
{\cal A}\/(g,\varphi)=\int{\left[\sqrt{|g|}\left(\varphi\tilde{R} -\frac{\omega_0}{\varphi}\varphi_i\varphi^i - U\/(\varphi) \right)+ {\cal L}_m\right]\,ds}
\end{equation}
where $\varphi\/$ is the scalar field, $\varphi_i := \de\varphi/de{x^i}\/$ and $U\/(\varphi)\/$ is the potential of $\varphi\/$. The matter Lagrangian ${\cal L}_m\/(g_{ij},\psi)\/$ is a function of the metric and some matter fields $\psi\/$. $\omega_0\/$ is the so called Brans--Dicke parameter \footnote{ However, it is worth noticing that this is a Brans--Dicke--like theory, not a proper Brans-Dicke theory, due to the presence of the self-interacting potential. This fact, in particular the presence of the potential, have to be stressed since misleading conclusions on the analogy between $f(R)$--gravity and Brans-Dicke theory could be drawn. For a recent discussion on this topic see \cite{arturoPLB}}.

The field equations, derived by varying with respect to the metric and the scalar field, are
\begin{equation}\label{0000.2}
\tilde{R}_{ij} -\frac{1}{2}\tilde{R}g_{ij}= \frac{1}{\varphi}\Sigma_{ij} + \frac{\omega_0}{\varphi^2}\left( \varphi_i\varphi_j  - \frac{1}{2}\varphi_h\varphi^h\/g_{ij} \right) 
+ \frac{1}{\varphi}\left( \tilde{\nabla}_{j}\varphi_i - \tilde{\nabla}_h\varphi^h\/g_{ij} \right) - \frac{U}{2\varphi}g_{ij} 
\end{equation} 
and
\begin{equation}\label{0000.3}
\frac{2\omega_0}{\varphi}\tilde{\nabla}_h\varphi^h + \tilde{R} - \frac{\omega_0}{\varphi^2}\varphi_h\varphi^h - U' =0
\end{equation} 
where $\Sigma_{ij}:= - \frac{1}{\sqrt{|g|}}\frac{\delta{\cal L}_m}{\delta g^{ij}}\/$ and $U' :=\frac{dU}{d\varphi}\/$.
Taking the trace of Eq. \eqref{0000.2} and using it to replace $\tilde R\/$ in eq. \eqref{0000.3}, one obtains the equation
\begin{equation}\label{0000.4}
\left( 2\omega_0 + 3 \right)\/\tilde{\nabla}_h\varphi^h = \Sigma + \varphi U' -2U
\end{equation} 
By a direct comparison, it is easy to see that for 
${\displaystyle \omega_0 =-\frac{3}{2}\/}$ and ${\displaystyle U\/(\varphi) =\frac{2}{\varphi}V\/(\varphi)\/}$ (where $V\/(\varphi)\/$ is defined in Eq. \eqref{2.8})  
eqs. \eqref{0000.2} become formally identical to the Einstein--like equations \eqref{2.7} 
for a metric--affine $f(R)\/$ theory. Moreover Eq. \eqref{0000.4} reduces to the algebraic equation
\begin{equation}\label{0000.5}
\Sigma + 2V'\/(\varphi) -\frac{6}{\varphi}V\/(\varphi) =0
\end{equation}   
relating the matter trace $\Sigma\/$ to the scalar field $\varphi\/$. 
In particular, it is  straightforward to verify that 
(under the condition $f''\not= 0\/$ \cite{CCSV1}) Eq. \eqref{0000.5} expresses  the inverse relation of Eq. \eqref{2.6}, namely
\begin{equation}\label{0000.6}
\Sigma + 2V'\/(\varphi) -\frac{6}{\varphi}V\/(\varphi) =0 \quad \Longleftrightarrow \quad \Sigma=F^{-1}\/((f')^{-1}\/(\varphi))
\end{equation}
being $F^{-1}\/(X) = f'\/(X)X - 2f\/(X)\/$. Metric--affine $f(R)\/$ theories (with torsion or {\it \`a
la\/} Palatini) are then dynamically equivalent to scalar--tensor theories with Brans--Dicke parameter ${\displaystyle \omega_0 =-\frac{3}{2}\/}$.

\section{The Cauchy problem of metric-affine $f(R)$-gravity in presence of  Klein-Gordon scalar field}

Let us consider now  the Cauchy problem for metric--affine 
$f(R)\/$-gravity where  a Klein-Gordon scalar field 
acting as a source. We shall derive sufficient conditions ensuring
the well-posedness of the problem. This result is 
obtained making use of the above stated dynamical equivalence 
between metric-affine $f(R)\/$-theories and $\omega_0=-\frac{3}{2}\/$ 
scalar--tensor theories (see also \cite{CV1,CV2,CV3}). 

Let us start by defining  a Klein-Gordon
scalar field $\psi$ whose dynamics is given by the self-interacting potential
${\displaystyle U(\psi)=\frac{1}{2} m^2\psi^2} $. The associated
stress-energy tensor is 
\begin{equation}\label{3.1}
\Sigma_{ij}= \de\psi/de{x^i}\de\psi/de{x^j}
-\frac{1}{2}g^{ij}\left(\de\psi/de{x^p}\de\psi/de{x^q}g^{pq} +
m^2\psi^2\right)
\end{equation}
The corresponding Klein-Gordon equation is given by
\begin{equation}\label{3.2}
\tilde\nabla_j\de\psi/de{x^i}g^{ij}=m^2\psi
\end{equation}
where $\tilde{\nabla}\/$ denotes the Levi-Civita covariant derivative induced by the metric $g_{ij}\/$. The trace of the tensor \eqref{3.1} is 
\begin{equation}\label{3.3}
\Sigma :=\Sigma_{ij}g^{ij} = -\de\psi/de{x^p}\de\psi/de{x^q}g^{pq} -2m^2\psi^2\,.
\end{equation}
Furthermore, let us consider a scalar--tensor theory, with Brans-Dicke parameter $\omega_0=-\frac{3}{2}\/$ and potential $U\/(\varphi)=\frac{2}{\varphi}V\/(\varphi)\/$, 
coupled with this  Klein-Gordon field. The corresponding field equations are the Einstein-like Eqs. \eqref{2.7}, Eq- \eqref{0000.5} relating the scalar field $\varphi\/$ to the trace $\Sigma\/$, and the Klein-Gordon Eq. \eqref{3.2}. In order to discuss  the Cauchy problem of such a theory, following \cite{CV1,CV3}, 
we begin by performing the conformal transformation $\bar{g}_{ij}=\varphi\/g_{ij}\/$. The Einstein-like Eqs. \eqref{2.7} assume then the simpler form (see for example \cite{CCSV1,Olmo})
\begin{equation}\label{3.4}
\bar{R}_{ij} - \frac{1}{2}\bar{R}\bar{g}_{ij} =
\frac{1}{\varphi}\Sigma_{ij} -
\frac{1}{\varphi^3}V\/(\varphi)\bar{g}_{ij}
\end{equation}
where $\bar{R}_{ij}\/$ and $\bar{R}\/$ are respectively the Ricci
tensor and the curvature scalar derived from the conformal metric
$\bar{g}_{ij}\/$. Similarly, a direct calculation shows that the Klein-Gordon equation, expressed in terms of the conformal metric $\bar{g}_{ij}\/$, becomes 
\begin{equation}\label{3.5}
-\de{\psi}/de{x^i}\bar{g}^{ij}\de{\varphi}/de{x^j} + \varphi\bar{\nabla}_j\de{\psi}/de{x^i}\bar{g}^{ij} = m^2\psi
\end{equation}
where $\bar{\nabla}_j\/$ denotes the covariant derivative associated with the conformal metric $\bar{g}_{ij}\/$. Also the trace $\Sigma\/$ can be expressed in function of $\bar{g}_{ij}\/$, that is 
\begin{equation}\label{3.6}
\Sigma=-\de\psi/de{x^p}\de\psi/de{x^q}\varphi\bar{g}^{pq} -2m^2\psi^2\,.
\end{equation}
Now, the relation \eqref{0000.5} links the scalar field $\varphi\/$ to the Klein--Gordon field $\psi\/$, its partial derivatives ${\displaystyle \frac{\partial\psi}{\partial x^i}}$ and the conformal metric $\bar{g}_{ij}\/$.  In Eqs. \eqref{3.4}, the quantity 
\begin{equation}\label{3.7}
T_{ij} := \frac{1}{\varphi}\Sigma_{ij} -
\frac{1}{\varphi^3}V\/(\varphi)\bar{g}_{ij}
\end{equation}
plays the role of the effective stress-energy tensor. Furthermore, it is worth noticing  that  conservation laws  can be related  in the Jordan and in the Einstein frame  \cite{CV1,CV3}, that is
\begin{Proposition}\label{Pro3.1}
Eqs. \eqref{2.7}, \eqref{2.8} and \eqref{0000.5} imply the usual conservation laws $\tilde{\nabla}^j\Sigma_{ij}=0\/$.
\end{Proposition}
\begin{Proposition}\label{Pro3.2}
The condition $\tilde{\nabla}^j\Sigma_{ij}=0\/$ is equivalent to the condition $\bar{\nabla}^jT_{ij}=0\/$.
\end{Proposition}
In the following discussion,  {\bf Proposition 2}  plays a crucial role. The Klein-Gordon Eq. \eqref{3.2} implies the conservation laws $\tilde{\nabla}^j\Sigma_{ij}=0\/$ 
and then the identity $\bar{\nabla}^jT_{ij}=0\/$. It turns out that the latter allows  to use harmonic coordinates to deal with the Cauchy problem, according to the  lines considered in \cite{yvonne,yvonne2,Wald}. 

We start by rewriting the Einstein--like Eqs. \eqref{3.4} in the equivalent form
\begin{equation}\label{3.8}
\bar{R}_{ij} = T_{ij} - \frac{1}{2}T\bar{g}_{ij}
\end{equation}
Assuming harmonic coordinates, i.e. local coordinates satisfying the conditions
\begin{equation}\label{3.9}
\bar{\nabla}_p\bar{\nabla}^p\/x^i= - \bar{g}^{pq}\bar{\Gamma}_{pq}^i =0\,,
\end{equation}
we can express Eqs. \eqref{3.8} as (see, for example, \cite{yvonne,Wald})
\begin{equation}\label{3.10}
\bar{g}^{pq}\frac{\partial^2 \bar{g}_{ij}}{\partial x^p \partial x^q} = f_{ij}\/(\bar{g},\partial\bar{g},\psi,\partial\psi)\,,
\end{equation}
where $f_{ij}\/$ indicate suitable functions depending  only on the metric $\bar{g}\/$, the scalar field $\psi\/$ and their first-order derivatives.

Moreover, from now on, we assume that Eq. \eqref{0000.5} is solvable with respect to the variable $\varphi\/$. In other words, we suppose to be able to derive from Eq. \eqref{0000.5} a function of the form
\begin{equation}\label{3.11}
\varphi=\varphi\/\left(\bar{g},\psi,\de\psi/de{x^p}\de\psi/de{x^q}\bar{g}^{pq}\right)
\end{equation}
expressing the scalar field $\varphi\/$ in terms of the metric $\bar{g}\/$, the Klein-Gordon field $\psi\/$ and its first order derivatives. In particular, from Eq. \eqref{3.6}, it is easily seen that the dependence of $\varphi\/$ on the derivatives of $\psi\/$ has to be  necessarily of the above form. The requirement on Eq. \eqref{0000.5} depends on the explicit form of the potential $V\/(\varphi)$. The latter is determined by the function $f(R)\/$ through the relation \eqref{2.8}. Therefore, the above assumption becomes a rule to select viable $f(R)$-models. In addition, from Eq. \eqref{3.11},  we derive the identity  
\begin{equation}\label{3.12}
\de\varphi/de{x^i}=\de\varphi/de{\left(\de\psi/de{x^s}\de\psi/de{x^t}\bar{g}^{st}\right)}2\de\psi/de{x^q}\bar{g}^{pq}\frac{\partial^2\psi}{\partial x^i \partial x^p} + f_i\/(\bar{g},\partial\bar{g},\psi,\partial\psi)\,.
\end{equation}
Inserting Eq. \eqref{3.12} in Eq. \eqref{3.5} and taking Eqs. \eqref{3.9} into account, we obtain the final form of the Klein--Gordon equation given by 
\begin{equation}\label{3.13}
\left(\bar{g}^{ip} -\frac{2}{\varphi}\de\varphi/de{\left(\de\psi/de{x^s}\de\psi/de{x^t}\bar{g}^{st}\right)}\de\psi/de{x^j}\bar{g}^{ji}\de\psi/de{x^q}\bar{g}^{pq}\right)\frac{\partial^2\psi}{\partial x^i \partial x^p} = f\/(\bar{g},\partial\bar{g},\psi,\partial\psi)
\end{equation}
In Eqs. \eqref{3.12} and \eqref{3.13}, $f_i\/$ and $f\/$ indicate suitable functions of $\bar{g}_{ij}\/$, $\psi\/$ and their first order derivatives only. 
Eqs. \eqref{3.10} and \eqref{3.13} describe a second order quasi-diagonal system of partial differential equations for the unknowns $\bar{g}_{ij}\/$ and $\psi\/$. 
The matrix of the principal parts of such a system is diagonal and its elements are the differential operators
\begin{subequations}\label{3.14}
\begin{equation}\label{3.14a}
\bar{g}^{pq}\frac{\partial^2}{\partial x^p \partial x^q}
\end{equation}
and
\begin{equation}\label{3.14b}
\left(\bar{g}^{ip} -\frac{2}{\varphi}\de\varphi/de{\left(\de\psi/de{x^s}\de\psi/de{x^t}\bar{g}^{st}\right)}\de\psi/de{x^j}\bar{g}^{ji}\de\psi/de{x^q}\bar{g}^{pq}\right)\frac{\partial^2}{\partial x^i \partial x^p}
\end{equation}
\end{subequations}
The operator \eqref{3.14a} is noting else but the wave-operator associated with the metric $\bar{g}_{ij}\/$, while the operator \eqref{3.14b} is very similar to the sound-wave-operator arising from the analysis of the Cauchy problem for General Relativity coupled with an irrotational perfect fluid \cite{yvonne,yvonne4}. To discuss the Cauchy problem for the system \eqref{3.10} and \eqref{3.13}, we can then follow the same arguments developed in \cite{yvonne,yvonne4}.
In particular, if the quadratic form associated with \eqref{3.14b} is of Lorentzian signature and if the characteristic cone of the operator \eqref{3.14b} is exterior to the metric cone, the system \eqref{3.10} and \eqref{3.13} is causal and  hyperbolic  in Leray sense \cite{Leray,yvonne3}. In such a circumstance, the corresponding Cauchy problem is well-posed in suitable Sobolev spaces. Still following \cite{yvonne,yvonne4}, if the signature of $\bar{g}_{ij}\/$ is $(+---)\/$, the required conditions are satisfied when the vector ${\displaystyle \de\psi/de{x^j}\bar{g}^{ij}\/}$ is timelike and 
the inequality
\begin{equation}\label{3.15}
-\frac{2}{\varphi}\de\varphi/de{\left(\de\psi/de{x^s}\de\psi/de{x^t}\bar{g}^{st}\right)}\geq 0
\end{equation}
holds. On the contrary, if the signature of $\bar{g}_{ij}\/$ is $(-+++)\/$, the inequality \eqref{3.15} has to be inverted. As already mentioned, the explicit expression of the function \eqref{3.11} depends on that of the potential \eqref{2.8} which is determined by the function $f(R)\/$. Therefore, the requirement \eqref{3.15} (or, equivalently, its opposite) can be a criterion to single out viable $f(R)\/$-models as shown in \cite{CV1}. An  example to illustrate the result is given in the next section.   

 \section{The  $f(R)= R+\alpha\/R^2$ case}
 
Let us consider the model $f\/(R)=R+\alpha\/R^2\/$. Taking into account that $F^{-1}\/(X)=f'\/(X) -2f\/(X)=-X\/$ and ${\displaystyle (f')^{-1}\/(\varphi)=\frac{\varphi -1}{2\alpha}\/}$, from the definition \eqref{2.8}, we get the identity
\begin{equation}\label{4.1}
V\/(\varphi)= \frac{1}{8\alpha}(\varphi -1)^2\varphi
\end{equation}
representing the effective potential for the considered model. Eq. \eqref{4.1} together with eqs. \eqref{0000.5} and \eqref{3.6} yield
\begin{equation}\label{4.2}
\varphi = \frac{\left( \frac{1}{2\alpha} + 2m^2\psi^2 \right)}{\left( \frac{1}{2\alpha} - \de\psi/de{x^s}\de\psi/de{x^t}\bar{g}^{st} \right)}
\end{equation}
that expresses the scalar field $\varphi\/$ as a function of the metric $\bar{g}_{ij}\/$, the Klein--Gordon field $\psi\/$ and its first order derivatives. Directly from \eqref{4.2} it follows
\begin{equation}\label{4.3}
-\frac{2}{\varphi}\de\varphi/de{\left(\de\psi/de{x^s}\de\psi/de{x^t}\bar{g}^{st}\right)}= - \frac{2}{\left( \frac{1}{2\alpha} - \de\psi/de{x^s}\de\psi/de{x^t}\bar{g}^{st} \right)}
\end{equation}
In the signature $(+---)\/$ for the metric $\bar{g}_{ij}\/$, it is immediately seen that the requirement \eqref{3.15} is automatically satisfied if $\alpha <0\/$ and if ${\displaystyle \bar{g}^{pq}\de\psi/de{x^q}\/}$ is a timelike vector field. 
If the signature is $(-+++)\/$, the condition becomes $\alpha >0\/$. It is worth noticing that the condition $\alpha <0\/$ ensures the well--posedness of the Cauchy problem for the  model $f\/(R)=R+\alpha\/R^2\/$ also when coupled with a perfect fluid \cite{CV1}.

\section{Conclusions}

According to the prescriptions given in \cite{yvonne,yvonne2,yvonne4}, we have formulated sufficient conditions to ensure the well--posedness of the Cauchy problem for metric-affine
$f(R)$-theories {\it \`a la\/} Palatini and with torsion, in
presence of Klein-Gordon scalar field acting as source.   As in the case of perfect fluid \cite{CV1}, the procedure allows
to select physically viable models. The presented results refute the criticisms advanced by some authors \cite{Faraoni} about the viability of metric-affine $f(R)$-theories.
The key points of the demonstration
are that the conservation laws  are preserved under the conformal transformation and the Bruhat arguments can
be applied if  suitable differential operators are defined for the Klein-Gordon field. 
The  $f(R)=R+\alpha R^2$ case is paradigmatic. The well-posedness strictly depends on the sign of the parameter $\alpha$ 
in connection with the kind of signature of the metric tensor.
As shown also in cosmological context, this can discriminate between physical and  unphysical models \cite{barrow}.

In conclusion,  the  Cauchy problem
can be, in general, well-formulated for $f(R)$-gravity in
metric-affine  as well as in metric formalism
\cite{Faraoni,CV3}. However, the well-posedness
strictly depends on the source and the parameters of the
theory. (e.g. $\alpha$  in the above
example). In the case of Klein-Gordon scalar field and perfect-fluid matter, it works,
essentially, because the problem can be reduced to the Einstein
frame by a conformal transformation and the Bruhat
arguments can be applied.

\end{document}